\documentclass[english,osajnl,9pt,twocolumn,twoside,showpacs]{revtex4-1}
\usepackage[T1]{fontenc}
\usepackage{float}
\usepackage{amsmath}
\usepackage{amssymb}
\usepackage{graphicx}

\makeatletter

\providecommand{\tabularnewline}{\\}

\usepackage{kotex}

\makeatother

\usepackage{babel}
\begin{document}
\title{  Circle-shaped Transformation Cavity via Center-Shift M\"{o}bius
Transfomation }
\author{\emph{Jinhang Cho}}
\affiliation{Digital Technology Research Center, Kyungpook National University,
Daegu 41566, Republic of Korea}
\author{\emph{Inbo Kim}}
\affiliation{Digital Technology Research Center, Kyungpook National University,
Daegu 41566, Republic of Korea}
\author{\emph{Sunghwan Rim}}
\affiliation{Digital Technology Research Center, Kyungpook National University,
Daegu 41566, Republic of Korea}
\author{\emph{Yong-Hun Lee}}
\affiliation{School of Electronics Engineering, Kyungpook National University,
Daegu 41566, Republic of Korea}
\author{\emph{Jung-Wan Ryu}}
\affiliation{Center for Theoretical Physics of Complex Systems, Institute for Basic
Science (IBS), Daejeon 34126, Republic of Korea}
\author{\emph{Muhan Choi}}
\email{mhchoi@ee.knu.ac.kr}

\affiliation{Digital Technology Research Center, Kyungpook National University,
Daegu 41566, Republic of Korea}
\begin{abstract}
Circle-shaped transformation cavities of shifted center are the most
simple transformation cavities in the sense that they have only spatial
index variation without boundary shape deformation. For these cavities,
we analyze the characteristics of conformal whispering gallery modes
according to center-shift coordinate transformation. For this study,
we use an alternative scheme for designing a transformation cavity
without using a shrinking parameter that is needed to satisfy total
internal reflection condition. It is observed that the isotropic emission
of whispering gallery modes in the uniform index circular cavity changes
to the bi-directional emission of conformal whispering gallery modes
as the center-shift increases.  Also, using a purity factor that
quantitatively measures in RV space the degree of intactness for
the internal wave pattern of resonant modes in transformation cavities
compared with corresponding whispering gallery mode in uniform disk
cavity, we show that the conformal whispering gallery modes  in the
circle-shaped transformation cavity has not only directionality but
also high persistence for characteristics of whispering gallery modes.
\end{abstract}
\pacs{xx.xx}
\maketitle

\section{Introduction}

The 2-dimensional (2D) dielectric microcavities have been intensively
studied over the last two decades due to the potential as ultra-small
high quality resonators needed in integrated optical circuits and
sensors \citep{Microcavity_R.K.Chang,MC_Vahala,MC_J.Wiersig_RMP}.
One of the core subjects in the studies is the whispering gallery
modes (WGMs), a very long-lived resonances trapped by the total internal
reflection (TIR). The ideal WGMs in the uniform refractive index circular
cavity can have very high quality-factor ($Q$-factor), but their
isotropic emission due to the rotational symmetry are serious drawback.
Many researches have been done to overcome this problem by deforming
the shape of the cavity or involving the scatterers, but it also
causes $Q$-factor degradation and undesired complex phenomena such
as wave chaos \citep{wave_chaos_stone,wave_chaos_An,wave_chaos_Shinohara,wave_chaos_Sunada}
and mode interaction effects \citep{EP_W.D.Heiss,ARC_J.Wiersig,ARC_annular_J.Wiersig,EP_Petermann_S.-Y.Lee,EP_Chiral_J.Wiersig,EP_Coupled_J.W.Ryu,EP_diff_m_C.W.Lee}.
As a novel approach to remedy above drawback, recently proposed are
transformation cavities (TCs), dielectric cavities with the gradient
refractive index (GRIN) designed based on conformal transformation
optics \citep{cWGM_Y.Kim_TC_Nature_Phot}. The anisotropic WGMs supported
in the TCs, so-called conformal whispering gallery modes (cWGMs),
provide directional tunneling emission as well as retain $Q$-factors
of corresponding WGMs in uniform disk cavity. The TCs can be designed
through various analytic functions of complex variables or the composite
functions of those. Also, even if an analytic form of conformal mapping
has singularities inside the cavity region or is difficult to be found
out for a given cavity shape, the numerical conformal mapping method
can be used for those cases \citep{S.J.Park_QCM_OE}.

The cWGMs in generic boundary deformed TCs have the property that
both effects of shape deformation and GRIN generated by conformal
mapping are intermingled. Circle-shaped TCs belong to a special class
TC that is ruled by only the effect of spatial refractive index variation,
completely free from the deformation effect.  The circle-shaped TC
can be formed by the center-shift conformal mapping which can be used
as an element of composite functions to achieve a realizable refractive
index profile or to break mirror symmetry for a given mirror symmetric
shaped TC. The investigation of the cWGM characteristics for the circle-shaped
TCs is also a groundwork on the researchs for various TCs using the
center-shift conformal mapping. 

This paper is organized as follows. We briefly review the center-shift
conformal transformation and the original method for designing TCs
using a shrinking parameter to support cWGMs in Sec. II. An alternative
scheme to construct the TCs without the shrinking parameter and a
circle-shaped TC constructed through this scheme is presented in Sec.
III. In Sec. IV, the characteristics of cWGMs in the circle-shaped
TC is numerically investigated and the purity factor for measuring
the intactness of cWGM is defined and discussed. Finally we give
a summary in Sec. V.

\section{Center-Shift Conformal Transformation and Circle-Shaped Transformation
Cavity with Shrinking Parameter}

\subsection{Center-Shift Conformal Transformation}

\begin{figure}[b]
\includegraphics[scale=0.15]{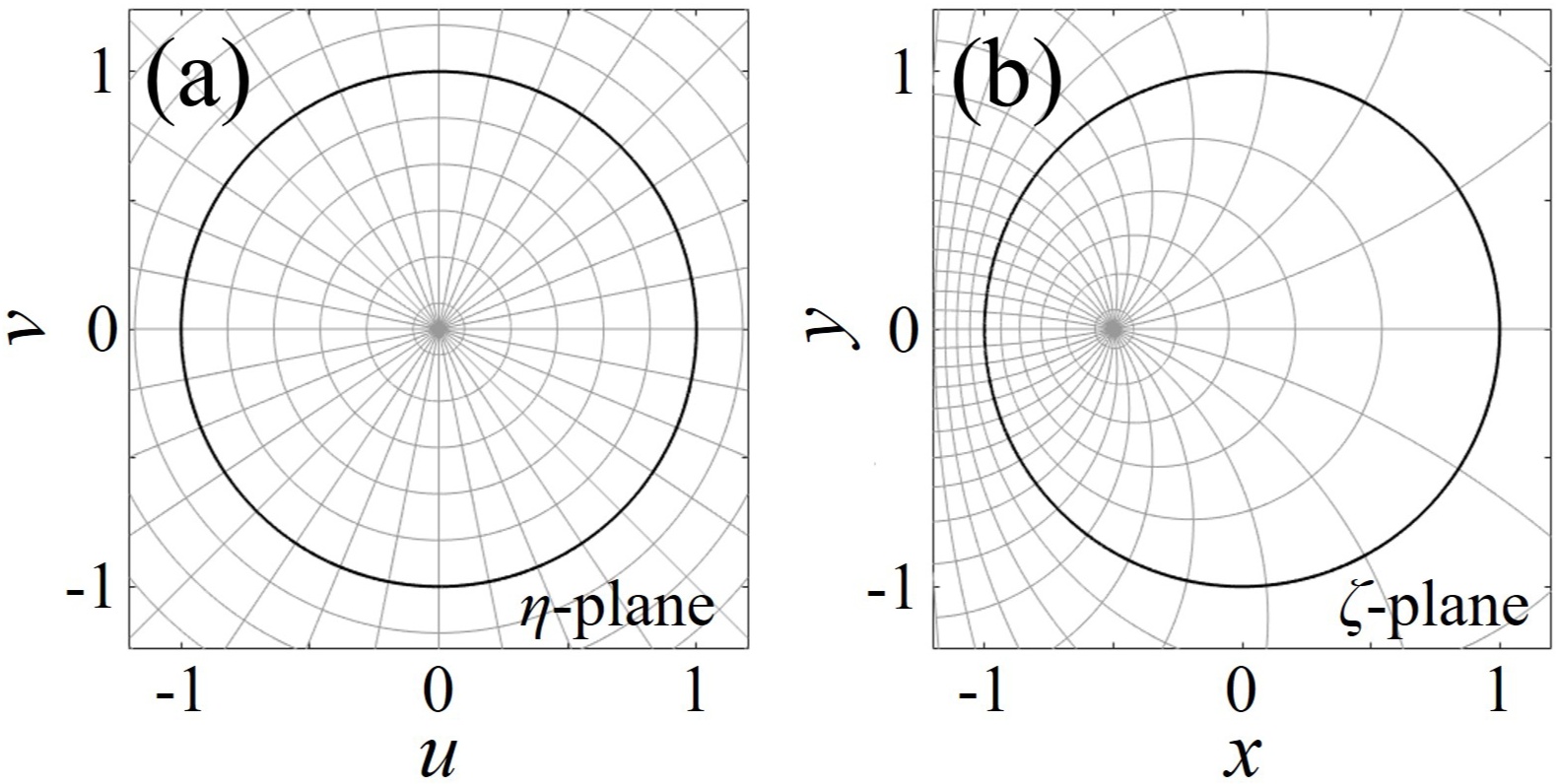}

\caption{Conceptual diagrams for center-shift conformal transformation from
(a) $\eta$-plane to (b) $\zeta$-plane. \label{fig:mobius_tr}}
\end{figure}

A mapping to transforms a unit circle in $\eta$-plane to a center-shifted
unit circle in $\zeta$-plane is given by the following form,

\begin{equation}
\zeta=\frac{\eta+\delta}{1+\delta^{*}\eta},\,\,\,\,\,\,\left|\delta\right|<1,\label{eq:CS_Map_complex}
\end{equation}
where $\eta=u+iv$ and $\zeta=x+iy$ are complex variables in the
complex planes, respectively, and $\delta$ is a complex valued parameter
representing the center-shift. This mapping forms a subgroup of M\"{o}bius
transformations and we will call it the center-shift conformal transformation
in this paper. Since the arbitrary center-shift can be described by
taking the line of shift as the $u$-axis ($\delta\in\mathbb{R}$),
so the above mapping can be simplified as follows:

\begin{equation}
\zeta=f(\eta)=\frac{\eta+\delta}{1+\delta\eta},\,\,\,\,\,\,\left|\delta\right|<1.\label{eq:CS_Map}
\end{equation}
 By the application of the above conformal mapping, a circle with
a radius of 1 in $\eta$-plane are transformed into a center-shifted
circle of the same size with the rotational symmetry broken in $\zeta$-plane
as shown in Fig. \ref{fig:mobius_tr}. For reference, the inverse
conformal mapping from $\zeta$-plane to $\eta$-plane is expressed
as follows:

\begin{equation}
\eta=g(\zeta)=\frac{\zeta-\delta}{1-\delta\zeta},\,\,\,\,\,\,\left|\delta\right|<1.\label{eq:inverse_CS_Map}
\end{equation}

\begin{figure*}
\includegraphics[width=18cm]{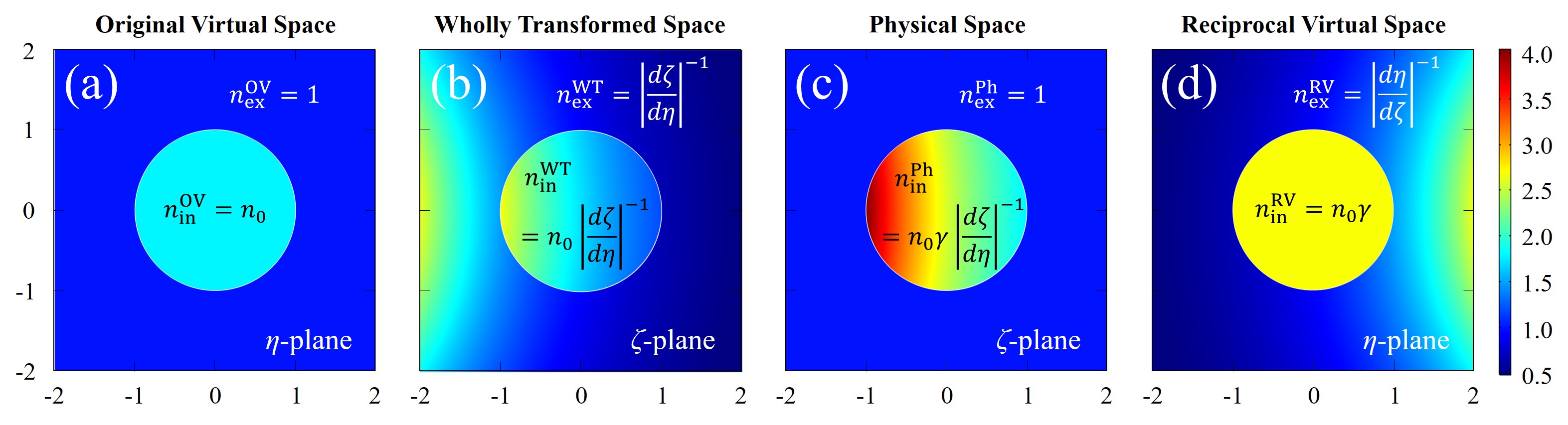}

\caption{Refractive index profiles in (a) original virtual, (b) wholly transformed,
(c) physical, and (d) reciprocal virtual spaces. These pictures were
drawn in circle-shaped transformation cavity without a shrinking parameter
at $n_{0}=1.8$, $\left|\delta\right|=0.2$, and $\gamma=\gamma_{\text{min}}$.
\label{fig:spaces}}
\end{figure*}

\subsection{Circle-Shaped Transformation Cavity with Shrinking Parameter}

Next, we look at the TCs which is previously proposed in \citep{cWGM_Y.Kim_TC_Nature_Phot}.
Considering the translational symmetry along the $z$-axis in a infinite
cylindrical dielectric cavity, the Maxwell equations can be reduced
to effective 2-dimensional scalar wave equation. Resonances in a TC
satisfying the outgoing-wave boundary condition, $\psi(\mathbf{r})\sim h(\phi,k)e^{ikr}/\sqrt{r}$
for $r\to\infty$, where position vector $\mathbf{r}=(x,y)=(r\cos\phi,r\sin\phi)$,
$k$ is the vacuum wavenumber, and $h(\phi,k)$ is the far-field angular
distribution of the emission, are obtained as the solutions of following
wave equation,

\begin{equation}
\left[\nabla^{2}+n^{2}(\mathbf{r})k^{2}\right]\psi(\mathbf{r})=0,\label{eq:HH}
\end{equation}
with the refractive index $n(\mathbf{r})$ given by 

\begin{equation}
n(\mathbf{r})=\begin{cases}
n_{0}\left|\frac{d\zeta}{d\eta}\right|^{-1}, & \text{(interior)}\\
1, & \text{(exterior)}
\end{cases}.\label{eq:RI_TC}
\end{equation}
For the transverse magnetic (TM) polarization, the wave function $\psi(\mathbf{r})$
represents $E_{z}$, the $z$ component of electric field, and both
the wave function $\psi(\mathbf{r})$ and its normal derivative $\partial_{v}\psi(\mathbf{r})$
are continuous across the cavity boundary. For the transverse electric
(TE) polarization, the wave function $\psi(\mathbf{r})$ represents
$H_{z}$, the $z$ component of magnetic field, and both $\psi(\mathbf{r})$
and $n(\mathbf{r})^{-2}\partial_{v}\psi$ are continuous across the
cavity boundary. By the outgoing-wave boundary condition, the resonances,
which have discrete complex wavenumbers $k_{r}$ with negative imaginary
parts, exponentially decay in time. The frequency and the lifetime
of a resonance are given by $\omega=c\text{Re}[k_{r}]$ and $\tau=-1/2c\text{Im}[k_{r}]$
where $c$ is light speed in vacuum, respectively. The quality factor
of a resonance is defined as $Q=2\pi\tau/T=-\text{Re}[k_{r}]/2\text{Im}[k_{r}]$
where the oscillation period of light wave is $T=2\pi/\omega$.

To describe the conventional method for constructing TCs supporting
cWGMs, four spaces are usually considered : original virtual (OV),
wholly transformed (WT), physical (Ph), and reciprocal virtual (RV)
spaces \citep{BEM_TC}. First, a disk cavity with a homogeneous refractive
index $n_{0}$ and a unit radius $R_{0}$ is considered in complex
$\eta$-plane called OV space. The uniform index disk cavity in the
OV space is conformally transformed to a cavity in complex $\zeta$-plane
called WT space through a entire spatial conformal mapping multiplied
by $\beta$, such as a resizable center-shift transformation,

\begin{equation}
\zeta=\beta f(\eta)=\beta\frac{\eta+\delta}{1+\delta\eta},\,\,\,\,\,\,\left|\delta\right|<1.\label{eq:CS_Map_beta}
\end{equation}

In the WT space, interior and exterior refractive index profiles of
the cavity are inhomogeneous but the relative refractive index at
the boundary interface remains homogeneous, because the two cavities
in OV and WT spaces are mathematically equivalent. Next, we set the
exterior GRIN profile in WT space to 1 considering the realistic physical
situation then, finally, we can obtain a circle-shaped TC in physical
space. The refractive index of the circle-shaped TC in physical space
can be derived from Eq. \eqref{eq:CS_Map_beta} as following form,

\begin{equation}
n(\mathbf{r})=\begin{cases}
n_{0}\left|\frac{(\beta-\delta\zeta)^{2}}{\beta(\delta^{2}-1)}\right|^{-1},\,\,\,\,\,\,\left|\delta\right|<1, & \text{(interior)}\\
1, & \text{(exterior)}
\end{cases}.\label{eq:RI_CSCircle_beta_TC}
\end{equation}
The relative refractive index at the boundary interface is not homogeneous
in the physical space. The heterogeneity of the relative refractive
index acts as an important factor in forming the resonance characteristics
which differ from in a uniform index disk cavity, and the reason can
be easily predicted through RV space which is mathematically equivalent
to the physical space. The RV space can be obtained from inverse conformal
transformation over the entire domain of physical space.

In order to obtain cWGMs in the TCs, we finally apply a specific value
$\beta_{\text{max}}$ or lower to $\beta$ such that the minimum value
of the internal refractive index profile is at least $n_{0}$. This
is a TIR minimum condition that reduces the size of the cavity and
at the same time allows the ray trajectory at the boundary to satisfy
the TIR angle. The TIR minimum condition in this case is obtained
as follows:

\begin{equation}
\beta_{\text{max}}\equiv\frac{1-\left|\delta\right|}{1+\left|\delta\right|},\,\,\,\,\,\,\left|\delta\right|<1.\label{eq:beta_max}
\end{equation}
 Here, we note that $\left|d\zeta/d\eta\right|^{-1}$ which forms
the refractive index of TCs, is a function of $\beta$, so not only
the cavity size but the GRIN profile changes according to the change
of $\beta$.

\section{Circle-Shaped Transformation Cavity without Shrinking Parameter}

In the aforementioned construncting method, the cavity maintains the
unit size in the RV space, but not in the physical space because of
the shrinking parameter. In particular, for the case of a circle-shaped
TC with no boundary shape change, it is easier to analyze pure spatial
index variation to completely exclude the parametrically changed variables
related to the boundary. Here, we propose a new TC design scheme that
maintains the dimensionless coordinates \citep{Microcavity_R.K.Chang}
in physical space by eliminating the size-scaling by confromal mapping.
We first consider a uniform index disk cavity in OV space ($\eta$-plane)
with the unit radius $R_{0}$ and the invariant reference refractive
index $n_{0}$ and then, conformally transform the entire space multiplied
by an index-proportion parameter $\gamma$ into a WT space ($\zeta$-plane)
through the mapping equation without $\beta$, Eq. \eqref{eq:CS_Map}.
Finally, forcing the exterior refractive index to 1 yields the TC
in physical space. As a result, the interior and exterior refractive
index in the physical space is as follows: 

\begin{equation}
n(\mathbf{r})=\begin{cases}
n_{\text{in}}^{{\scriptscriptstyle \text{Ph}}}=n_{\text{0}}\gamma\left|\frac{d\zeta}{d\eta}\right|^{-1}\,, & \text{(interior)}\\
n_{\text{ex}}^{{\scriptscriptstyle \text{Ph}}}=1\,, & \text{(exterior)}
\end{cases}.\label{eq:RI_DTC}
\end{equation}
Additionally, through the inverse transformation  from $\zeta$-plane
to $\eta$-plane, the interior and exterior refractive index in RV
space can be obtained as follows:

\begin{equation}
\tilde{n}(\tilde{\mathbf{r}})=\begin{cases}
n_{\text{in}}^{{\scriptscriptstyle \text{RV}}}=n_{\text{0}}\gamma\,, & \text{(interior)}\\
n_{\text{ex}}^{{\scriptscriptstyle \text{RV}}}=\left|\frac{d\eta}{d\zeta}\right|^{-1}\,, & \text{(exterior)}
\end{cases}\label{eq:RI_DTC_RV}
\end{equation}
where position vector $\tilde{\mathbf{r}}=(u,v)$. We shortly call
this newly constructed TC to $\gamma$-type TC and, for clarity, we
refer to the conventional TC as the $\beta$-type TC. $\gamma$-type
TC can satisfy the TIR condition by increasing the interior refractive
index in physical space independently of the profile-generating factor,
$\left|d\zeta/d\eta\right|^{-1}$ of Eq. \eqref{eq:RI_DTC}, without
reducing the cavity size. This is the most noticeable difference from
the $\beta$-type TC, in which decreasing $\beta$ reduces the cavity
size and simultaneously increases the profile-generating factor of
Eq. \eqref{eq:RI_TC} as a whole. To help understand, we have drawn
conceptual diagrams of four spaces using in the $\gamma$-type circle-shaped
TC in Fig. \ref{fig:spaces}.

$\beta$-type and $\gamma$-type are perfectly identical systems in
the viewpoint of physics. In the case of $\gamma$-type, $k_{r}$
means the dimensionless resonant wavenumber and multiplying $k_{r}$
by $n_{0}\gamma$ changed by TIR condition gives the internal dimensionless
resonant wavenumber, $\kappa\equiv n_{0}\gamma k_{r}R_{0}$. The free-space
wavenumber in $\beta$-type is obtained by dividing $k_{r}$ in $\gamma$-type
by $\beta$. One can properly select and use at convenience among
the two types. $\gamma$-type can be more advantageous in terms of
theoretical and numerical study such as analyzing the changes of resonance
distribution in a given wavelength region, adjusting the target frequency
of the resonance to be observed, tracing a specific resonance to finding
the optimizing conditions, or reobtaining the refractive index profile
for designing on real fabrication.

\begin{table}[H]
\begin{centering}
\begin{tabular}{|l|c|c|}
\hline 
 & $\beta$-type TC & $\gamma$-type TC\tabularnewline
\hline 
\hline 
conformal transformation & $\zeta=\beta f(\eta)$ & $\zeta=f(\eta)$\tabularnewline
\hline 
refractive index in RV space & $n_{\text{in}}^{{\scriptscriptstyle \text{RV}}}=n_{0}$ & $n_{\text{in}}^{{\scriptscriptstyle \text{RV}}}=n_{0}\gamma$\tabularnewline
\hline 
radius of cavity in physical space & $\beta R_{0}=\beta$ & $R_{0}=1$\tabularnewline
\hline 
refractive index in physical space & $n_{\text{in}}^{{\scriptscriptstyle \text{Ph}}}=n(\zeta,\beta)$ & $n_{\text{in}}^{{\scriptscriptstyle \text{Ph}}}=n(\zeta)\gamma$\tabularnewline
\hline 
\end{tabular}
\par\end{centering}
\caption{Comparison between $\beta$-type and $\gamma$-type TCs \label{tab:diff_TC_DTC}}
\end{table}

\begin{figure}
\includegraphics[scale=0.3]{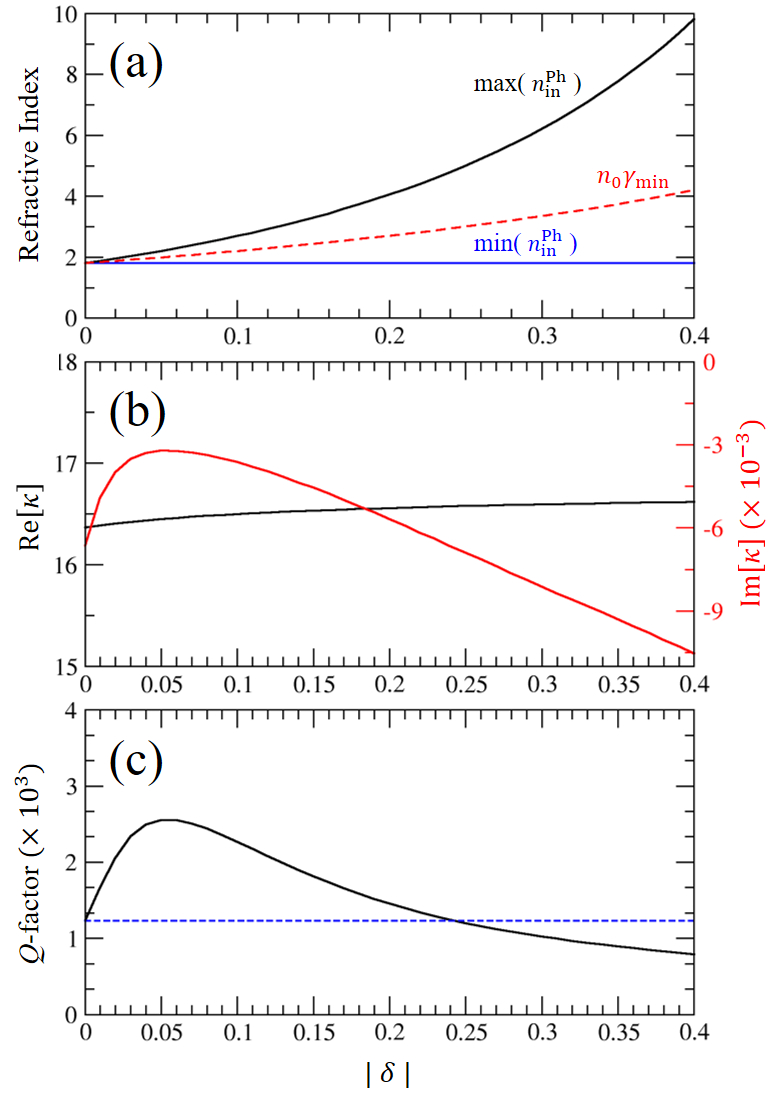}

\caption{(a) is the changes of the range of refractive index profile $n_{\text{in}}^{{\scriptscriptstyle \text{Ph}}}$
(black and blue solid line) and the value of $n_{0}\gamma_{\text{min}}$
(red dashed line) according to the center-shift parameters $\left|\delta\right|$
in the $\gamma$-type circle-shaped TC with $n_{0}=1.8$ and $\gamma=\gamma_{\text{min}}$.
(b) and (c) are the changes of internal dimensionless wavenumber $\kappa$
and $Q$-factor for M(13,1) that varies with $\left|\delta\right|$,
respectively. Black and red lines in (b) are the real and the imaginary
parts of $\kappa=n_{0}\gamma k_{r}R_{0}$, respectively. Blue dashed
lines in (c) are $Q$-factor in the uniform index disk cavity with
$n_{0}$. \label{fig:delta_vs_n_Q}}
\end{figure}

By the center-shift conformal transformation Eq. \eqref{eq:CS_Map},
the disk cavity with homogeneous interior refractive index $n_{0}$
in OV space is transformed to the $\gamma$-type circle-shaped TC
with the following inhomogeneous interior refractive index profile
in the physical space.  

\begin{equation}
n_{\text{in}}^{{\scriptscriptstyle \text{Ph}}}(\zeta)=n_{\text{0}}\gamma\left|\frac{(1-\delta\zeta)^{2}}{(\delta^{2}-1)}\right|^{-1},\,\,\,\,\,\,\left|\delta\right|<1,\label{eq:RI_CSCircle_in}
\end{equation}
 Also, we can derive the profile of exterior refractive index for
$\gamma$-type TC in the RV space through the inverse conformal mapping
Eq. \eqref{eq:inverse_CS_Map} as follows: 

\begin{equation}
n_{\text{ex}}^{{\scriptscriptstyle \text{RV}}}(\eta)=\left|\frac{(1-\delta^{2})}{(1+\delta\eta)^{2}}\right|,\,\,\,\,\,\,\left|\delta\right|<1.\label{eq:RI_CSCircle_RV_ex}
\end{equation}
 To obtain cWGMs formed by TIR, the condition is required that the
minimum value of the interior refractive index profile is at least
greater than $n_{0}$ and we define $\gamma$ satisfying this condition
as $\gamma_{\text{min}}$. In the case of $\gamma$-type circle-shaped
TC, $n_{\text{in}}^{{\scriptscriptstyle \text{TP}}}$ has the minimum
value at $\eta=-\delta/\left|\delta\right|$, thus $\gamma_{\text{min}}$
is given as follows:    

\begin{equation}
\gamma_{\text{min}}\equiv\frac{1+\left|\delta\right|}{1-\left|\delta\right|},\,\,\,\,\,\,\left|\delta\right|<1.\label{eq:gamma_min}
\end{equation}
Comparing the above equation with Eq. \eqref{eq:beta_max} used in
$\beta$-type TC, we can see that $\gamma_{\text{min}}$ lies in the
relationship of $1/\beta_{\text{max}}$.

\begin{figure*}
\includegraphics[width=17.9cm]{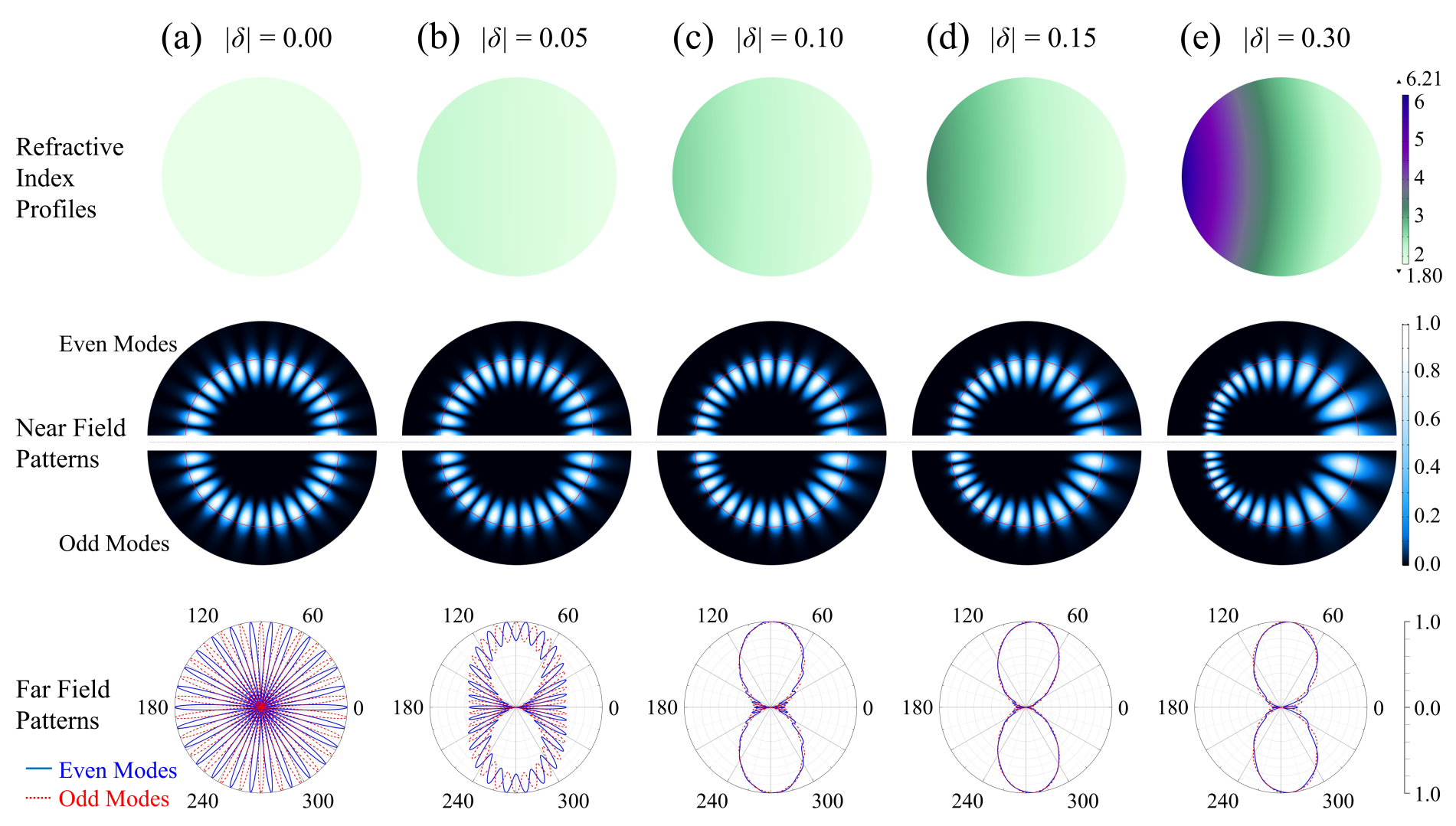}\caption{Refractive index profile and near field and far field intensity patterns
for the mode pair of M(13,1) at the shifting parameter (a) $\left|\delta\right|=0$,
(b) $0.05$, (c) $0.10$, (d) $0.15$, and (e) $0.30$ in the $\gamma$-type
circle-shaped TC with $n_{0}=1.8$ and $\gamma=\gamma_{\text{min}}$.
The refractive index of outside cavity set to air. In each step, up-side
(down-side) of near field pattern and blue solid (red dashed) line
in far-field pattern are the even-parity (odd-parity) mode. \label{fig:RI =000026 patterns}}

\end{figure*}

\section{Numerical Results}

\subsection{cWGMs in Circle-Shaped Transformation Cavity}

Using the $\gamma$-type circle-shaped TC presented above, we investigate
the characteristic changes of a specific cWGM for TM polarization
by changing in the range of $0\leq\left|\delta\right|\leq0.4$ under
the fixed conditions of reference refractive index $n_{0}=1.8$ and
index-proportion parameter $\gamma=\gamma_{\text{min}}$. These results
can be obtained from either finite element method (FEM) or boundary
element method (BEM) \citep{BEM_TC}, FEM-based COMSOL Multiphysics
were used in this study. Also, in this paper, we assign cWGMs to M($m$,
$l$) as a combination of mode indices corresponding to the angular
momentum index $m$ and the radial nodal number $l$ in the uniform
index disk cavity. According to the change of $\left|\delta\right|$,
the interior refractive index in OV space and the maximum and minimum
value of interior refractive index profile in physical spaces are
changed as shown in Fig. \ref{fig:delta_vs_n_Q} (a). \emph{ }As
$\left|\delta\right|$ increases, the maximum value of refractive
index profile $n_{\text{in}}^{{\scriptscriptstyle \text{Ph}}}$ gradually
widens the overall variation of the index profile with  $n_{0}$
as the baseline. At the same time, $\gamma_{\text{min}}$ also increases
to satisfy the TIR condition as shown in Fig. \ref{fig:delta_vs_n_Q}
(b).

The internal dimensionless wavenumbers $\chi$ and $Q$-factors for
the mode pair of M(13,1) are changed as shown in Fig. \ref{fig:delta_vs_n_Q}
(c) and (d), respectively. In the uniform index disk cavity case with
$\left|\delta\right|=0$, all resonances except for the case with
$m=0$ are in doubly degenerate states due to the rotational symmetry.
Each degenerate pair is nearly degenerate under the condition of $\left|\delta\right|>0$
and each nearly degenerate pair can be divided into even- and odd-parity
modes due to the mirror symmetry for the $x$-axis. Nevertheless,
in the range we show, the wavelenth values for the cWGM pair appear
to overlap almost one with no significant deviation. In terms of the
RV space shown in Fig. \ref{fig:spaces} (d)), it means that the cWGM
pair is very little affected by the pure change in relative refractive
index at the boundary interface caused by $\left|\delta\right|$ in
our range.

\emph{ }As shown in Fig. \ref{fig:delta_vs_n_Q} (c), $\text{Re}[\kappa]$
associated with the internal wavelength of the mode pair of M(13,1)
does not change significantly with the varying in $\left|\delta\right|$,
on the other hand, $\text{Im}[\kappa]$ grows larger than that in
the homogeneous case and then decreases from a certain threshold (in
our case, about $\left|\delta\right|=0.05$) as $\left|\delta\right|$
increases. It is directly reflected in the $Q$-factor in Fig. \ref{fig:delta_vs_n_Q}
(d). The temporary rise of $Q$-factor, which is typical aspect of
cWGMs in TC satisfying the TIR condition \citep{cWGM_Y.Kim_TC_Nature_Phot,optimi_limacon_J.-W.Ryu},
is caused by the overall increasement in the refractive index profile
of TC due to the TIR condition, and the increament of $\gamma_{\text{min}}$
in the $\gamma$-type TC well explains why such behavior occurs.

\emph{ }For the case of $\ensuremath{\left|\delta\right|\ne0}$,
the relative refractive index at the boundary interface and the interior
refractive index profile exhibit a dipole distribution similar to
the case of the lima\c{c}on TC \citep{cWGM_Y.Kim_TC_Nature_Phot}.
Considering the emitting mechanism described through the Husimi function
\citep{Husimi_I.Kim_OE}, cWGMs in the circle-shaped TC can be predicted
to have a similar mode properties. We show the stepwise change of
refractive index profile and near- and far-field intensity patterns
for a even-odd mode pair of M(13,1) at $\left|\delta\right|=0$, $0.05$,
$0.10$, $0.15$, and $0.30$ in Fig. \ref{fig:RI =000026 patterns}.
\emph{ }As $\ensuremath{\left|\delta\right|}$ increases, the dipole
distribution of the refractive index becomes more pronounced due to
the increase in the variation width of the index profile, while the
near field intensity pattern at all steps maintains a cWGM morphology
confined well along the boundary.

In Ref. \citep{Husimi_I.Kim_OE,optimi_limacon_J.-W.Ryu}, it has already
discussed that the emission of cWGMs satisfying the TIR condition
is tunneled out as an evanescent wave in the region where the refractive
index is relatively low and, as the deformation parameter increases,
the nearly flat intensity band structure on the Husimi function for
cWGMs is almost unchanged, while the shape of the critical line is
further bent by the variation of the relative refractive index at
the boundary interface. Such non-constant critical angle creates a
unique light emission mechanism which the light tunnels out at regions
where the band structure is relatively close to the critical angle.
i.e., where the relative refractive index at the boundary interface
is relatively low.

In the TCs with a dipole distributed refractive index profile, the
critical line approaches the band structure only in one place, which
creates a single point emitting mechanism, and their external waves
form bi-directional emission if they have an axis symmetry. We can
confirm it through the far field intensity patterns in Fig. \ref{fig:RI =000026 patterns}.
\emph{ }As $\left|\delta\right|$ increases, the isotropic emission
of mode pair when $\left|\delta\right|=0$ turns into the bi-directional
emission and the mode pairs in each step have the same tendency for
the far-field intensity distribution regardless of parity. The bi-directionality
is bestly improved at $\left|\delta\right|=0.15$. Incidentally, if
$\left|\delta\right|\geqq0.2$, the maximum value of refractive index
profile requires a large rise above 4 which is difficult to implement,
but the far field distribution still has bi-directionality.

\begin{figure*}
\includegraphics[scale=0.45]{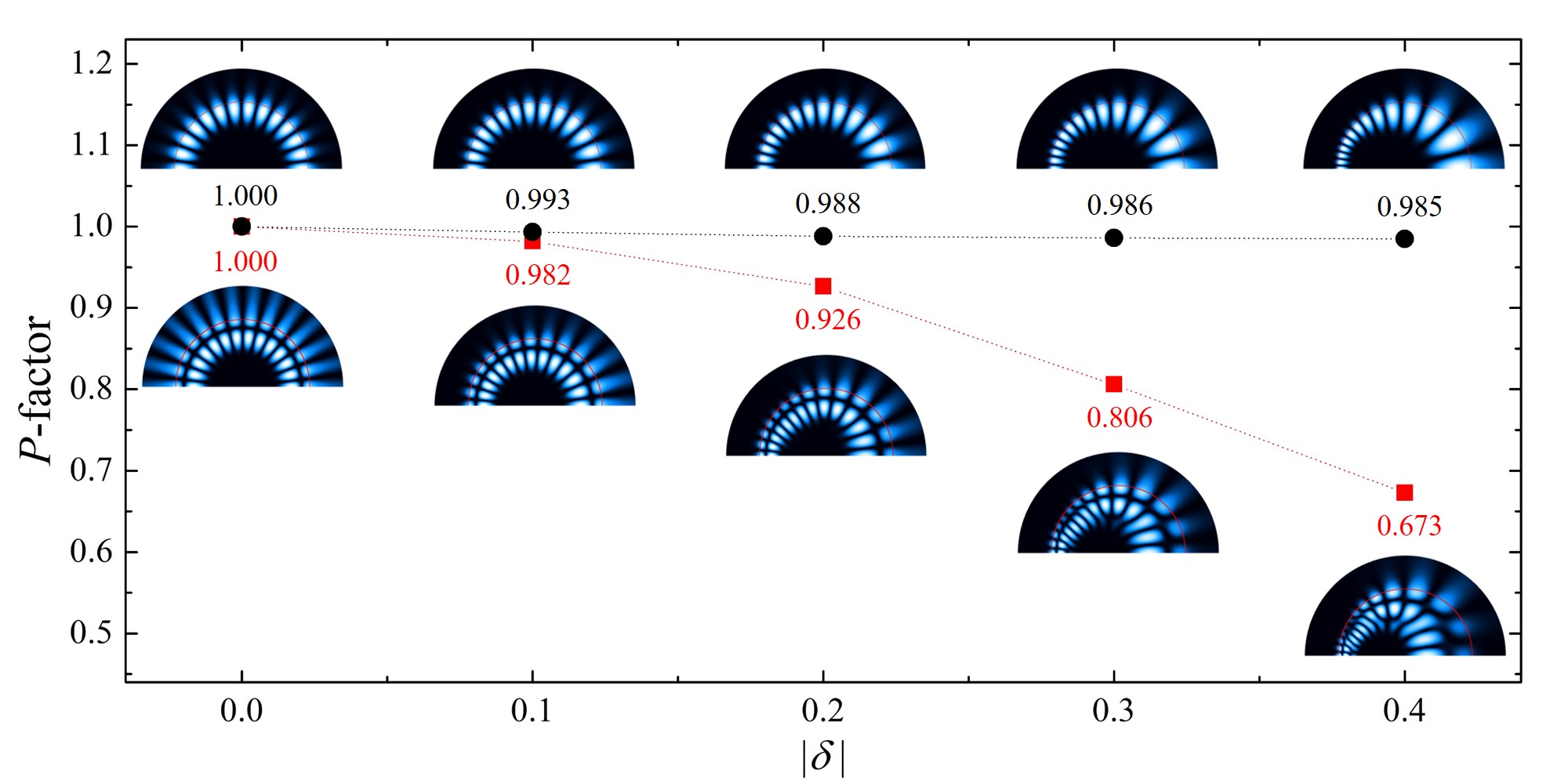} 

\caption{The change of $P$-factors for even-parity modes of (a) M(13,1) and
(b) M(13,2) according to $\left|\delta\right|$ in $\gamma$-type
circle-shaped TC with $n_{0}=1.8$ and $\gamma=\gamma_{\text{min}}$.
The data pickup domain for $P$-factors is set to $r_{{\scriptscriptstyle \text{RV}}}=0.8$
and the intensity of the near field patterns is normalized linear
scale. \label{fig:purity_1D}}

\end{figure*}

\subsection{Purity Factor for cWGMs}

We are here to present another tool for characterizing cWGMs. RV space
is easier to analyze the mode characteristics since it is the equvalent
space with physical space and formed with unit disk cavity with uniform
refractive index. In general, when the uniform index disk cavity is
progressively distorted by the shape deformation, WGMs with perfect
rotational symmetry begin to lose their inherent properties, accompained
by the $Q$-spoiling and the pattern distortion by the synthesis of
several angular momentum components. The angular memontum decomposition
is very useful method for analyzing such the angular momentum distribution
\citep{AMD}. The spread of $m$ derived from the analysis of angular
momentum components for a resonant mode can be used to gauge how much
the resonance in a slightly deformed cavity is distorted from a specific
resonance in the uniform index disk cavity. Such distortion also occurs
at the resonances in TCs. In the case of homogeneous cavities, the
distortion is due to the effect of shape deformation, whereas in the
case of circle-shaped TC without shape deformation, it is caused by
the non-uniformity of the relative refractive index at the boundary
interface. The variation of the distortion rate for each mode according
to the change of the system parameters has different criteria according
to the lifetime and wavelength of the resonance, but it is sufficient
to check how long the WGM characteristics of the uniform index disk
cavity are maintained in the cWGM.

To analyze the inherent mode properties inside TCs, we introduce the
angular momentum distribution in RV space equivalent with physical
space. In RV space, the wave function inside the uniform index disk
cavity can be expanded to cylindrical harmonics in polar coordinates
as follows: 

\begin{equation}
\psi(r_{{\scriptscriptstyle \text{RV}}},\phi_{{\scriptscriptstyle \text{RV}}})=\sum_{m=-\infty}^{\infty}\alpha_{m}J_{m}\left(\kappa\frac{r_{{\scriptscriptstyle \text{RV}}}}{R_{0}}\right)e^{im\phi_{{\scriptscriptstyle \text{RV}}}},\label{eq:AMD_RV}
\end{equation}
where $r_{{\scriptscriptstyle \text{RV}}}(<R_{0})$ and $\phi_{{\scriptscriptstyle \text{RV}}}$
are the radius and angle of the position on the data pickup domain,
respectively, + (-) signs of angular momentum index mean CCW (CW)
traveling-wave components, $\alpha_{m}$ is the angular momentum distribution
for $m$, and $J_{m}$is the $m$th-order Bessel function of the first
kind. $\alpha_{m}$ is obtained through the Fourier expansion of the
above equation. Note that the data pickup domain must be chosen as
a concentric circle with a center of mass in the RV space, taking
into account the spatial transformation by conformal mapping. From
the above angular momentum distribution, our newly proposed measurand
that estimates the mode distortion rate, namely purity factor, can
be simply defined as follows :

\begin{equation}
P=\frac{\left|\alpha_{d}\right|^{2}}{\sum_{m=0}^{\infty}\left|\alpha_{m}\right|^{2}}\,,\label{eq:P-factor}
\end{equation}
where $\alpha_{d}$ is that for a dominant angular momentum index
$d$. Under the our situation of standing wave conditions by axis
symmetry, we only need to take one component, CCW or CW. This $P$-factor
means the contribution of resonance with $m=d$ in the uniform index
disk cavity in forming a specific resonance we are going to observe
in TCs.

To investigating the change of distortion rate for the cWGM, we obtained
the above $P$-factor for M(13,1) in the range of $0\leq\left|\delta\right|\leq0.4$
where the data pickup domain $r_{\text{RV}}=0.8$. We plotted it in
Fig. \ref{fig:purity_1D} attaching the $P$-factor for M(13,2) for
comparison. The $\left|\delta\right|$-dependent increase in the heterogeneity
of the relative refractive index at the boundary interface, which
acts as an effective deformation effect, reduces overall the $P$-factor
for both modes. Here, it should be noted that the variation of the
$P$-factor for M(13,1) is very small. It means that the resonance
properties in the uniform index disk cavity are maintained fairly
well in cWGMs with $l=1$. Whereas M(13,2), which relatively less
satisfies the TIR condition, reacts more sensitively to the center-shift
parametric changes on the boundary relative refractive index variation,
resulting in a greater collapse of the $P$-factor.

\section{Summary}

In this paper, we have newly proposed a construction scheme for TC
without the shrinking parameter related to TIR condition, and, in
the circle-shaped TC using it, investigated the characteristic changes
of a cWGM of which isotropic emission breaks in bi-directional emission
as the center-shift parameter increases. The enhancement of distortion
effect on the modes due to the pure spatial refractive index variation
according to the center-shift parameter can be verified through the
newly defined purity factor $P$ indicating how much the nature of
WGM in the uniform index disk cavity is maintained via the angular
momentum distribution in the RV space. In conclusion, it has been
examined that the circle-shaped TC can produce the bi-directional
emitting fundamental cWGMs of which $P$-factor is nearly one.

\section*{Acknowledgments}

We would like to thank Y. Kim and S.-J. Park for helpful discussions.
This work was supported by the National Research Foundation of Korea
(NRF) grant funded by the Korean government (MSIP) (2017R1A2B4012045
and 2017R1A4A1015565) and the Institute for Basic Science of Korea
(IBS-R024-D1).

\end{document}